\documentclass[twocolumn,showpacs,preprintnumbers,amsmath,amssymb,floatfix,superscriptaddress]{revtex4}

\usepackage{epsfig}
\usepackage{graphicx}% Include figure files
\usepackage{dcolumn}% Align table columns on decimal point
\usepackage{bm}% bold math

\newcommand{\be}{\begin{equation}}
\newcommand{\ee}{\end{equation}}
\newcommand{\ua}{\uparrow}
\newcommand{\da}{\downarrow}

\newcommand{\ox}{\omega_x}
\newcommand{\oy}{\omega_y}
\newcommand{\oz}{\omega_z}
\newcommand{\op}{\omega_\perp}

\begin{document}

%\preprint{APS/123-QED}

\title{Destroying superfluidity by rotating a Fermi gas at Unitarity}

\author{I. Bausmerth}
\author{A. Recati}\email{recati@science.unitn.it}
\author{S. Stringari}
\affiliation{Dipartimento di
Fisica, Universit\`a di Trento and CNR-INFM BEC Center, I-38050
Povo, Trento, Italy}

\begin{abstract}
We study the effect of the rotation on a harmonically trapped Fermi gas at zero temperature under the assumption that vortices are not formed. We show that at unitarity the rotation produces a phase separation between a non rotating superfluid (S) core and a rigidly rotating normal (N) gas. The interface between the two phases is characterized by a density discontinuity $n_{\rm N}/n_{\rm S}= 0.85$, independent of the angular velocity. The depletion of the superfluid  and the angular momentum of the rotating configuration are calculated as a function of the angular velocity. The conditions of stability are also discussed and the critical angular velocity for the onset of a spontaneous quadrupole deformation of the interface is  evaluated.
%The conditions of energetic and dynamic stability  are also discussed. 
\end{abstract}
%\pacs{Valid PACS appear here}

\maketitle

The effect of the rotation on the behavior of a superfluid is a longstanding subject of investigation in condensed matter as well as in nuclear systems \cite{Donnelly, Broglia}. Due to the irrotationality constraint imposed by the existence of the order parameter a superfluid cannot rotate like a normal fluid. This implies the quenching of the moment of inertia at small angular velocities $\Omega$ and the occurrence of quantized vortices at higher $\Omega$. These peculiar features have been experimentally studied in a systematic way in superfluid helium and, more recently, in Bose-Einstein condensates (see e.g. \cite{Cornell}).

The recent  realization of ultracold  Fermi gases with tunable interaction close to a Feshbach resonance  has opened new opportunities to investigate the  effects of superfluidity along the so called BCS-BEC crossover (for recent reviews see, for example, \cite{rmpwilly,rmpstefano}). While on the BEC side of the resonance (small and positive values of the scattering length) the physics of a Fermi superfluid approaches the behavior of a Bose-Einstein condensed gas of dimers, on the opposite BCS side (small and negative values of the scattering length) superfluidity is much more fragile because of the weakness  of Cooper pairs. A further challenging regime is given by the unitary gas of infinite scattering length where the system exhibits a universal behavior and, in many respects,   behaves like a strongly interacting fluid. Quantized vortices have been recently observed on both sides of the resonance \cite{MIT1}, including unitarity, providing a unifying picture of the  behavior of the system along the crossover. 
  
Previous theoretical work on superfluid rotating Fermi gases has mainly focused on the dynamics \cite{Mauro} and on the instability \cite{Zhai,Cooper} of  configurations with high vortex density.
The purpose of this Letter is to investigate the behavior of the gas {\itshape{when quantized vortices are not formed}}.
Experimentally this scenario is achievable through an adiabatic ramping of the the angular velocity of the rotating trap starting from a configuration at rest. In the case of trapped BEC's this procedure has permitted to reach values of angular velocity significantly higher than the Feynman critical angular velocity for the formation of quantized vortices whose nucleation is inhibited by the presence of a barrier. 
{\begin{figure}[htb]
\begin{center}
\includegraphics[height=7cm] {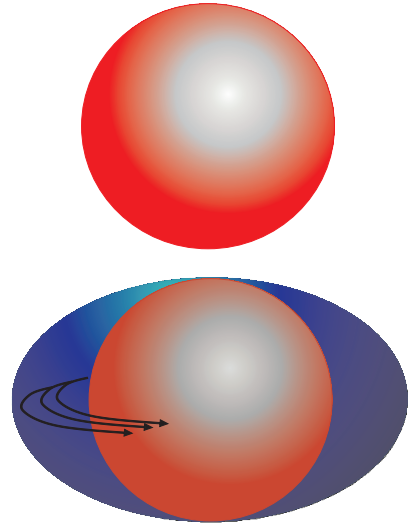}
\end{center}
\caption{Upper panel: without rotation the system is fully superfluid. Lower panel: the  rotation of the trap favours the formation of a phase separated state, with a  superfluid core surrounded by a normal part exhibiting the bulge effect of the rotation ($\Omega=0.45\op$).}
\label{fig:fancy}
\end{figure}}
Under these conditions new physical phenomena  occur like, for example, the spontaneous breaking of rotational symmetry caused by the energetic instability of the surface modes \cite{alessio} and, at even higher angular velocities, the occurrence of dynamic instabilities \cite{castin}. These phenomena have been observed experimentally in BEC's \cite{Dalibard1}, confirming in a qualitative and quantitative way the correctness of the irrotational hydrodynamic picture of rotating superfluids. 

The key point of the present work is that trapped superfluid Fermi gases  can behave quite differently from BEC's because  pairs are easily broken by the rotation. In the following we will restrict our discussion to the most relevant unitary gas. While at small angular velocities the superfluid is unaffected by the rotation of the trap \cite{Cozzini}, we predict that at higher angular velocities the rotation results in a phase separation between a non rotating superfluid core and a rigidly rotating normal component (see Fig.(\ref{fig:fancy})). The mechanism of pair breaking is very intuitive. In fact near the border of the cloud, where the density $n$ is small, the energy cost for destroying superfluidity is also small, being proportional to $n^{2/3}$. Viceversa the centrifugal energy gained by the normal phase can be large, being proportional to $\Omega^2R^2$ where $R$ is the radius of the cloud. Notice that this mechanism cannot occur in the BEC regime due to the high energy cost needed to break the dimers. Furthermore, differently from the nucleation of vortices, the formation of the normal part is not inhibited by the presence of a barrier as all the relevant energy scales are vanishingly small near the border and  the  energy gain is ensured as soon as one starts rotating the trap.
The appearance of a normal part due to rotation is not peculiar of ultra-cold atomic Fermi gases. For example, superfluidity is known to be unstable in nuclei rotating  with high angular velocity \cite{Broglia}. Such an effect is also related to the so-called intermediate state in type-I superconductors, where the role of the rotation is played by the external magnetic field (see, e.g., \cite{degennes}).

We are interested in a Fermi gas confined by a harmonic  potential 
%$V({\bf x})=m(\op^2 (1+\delta)x^2+\op^2 (1-\delta) y^2+\oz^2 z^2)$
$V({\bf r})=m(\ox^2 x^2+\oy^2 y^2+\oz^2 z^2)/2$, rotating with angular velocity $\Omega$ along $z$. We study the problem in the frame rotating with the trap where the potential is static  and the Hamiltonian is ${H-\Omega L_{\rm Z}}$.
In the local density approximation the energy of the rotating configuration at zero temperature can be written as 
\be
E=\int d{\bf r}\left\lbrack\epsilon(n)+V({\bf r})+\frac{1}{2}mv^2-m\Omega({\bf r}\times{\bf v})_{\rm Z}-\mu\right\rbrack n\ ,
\label{eq:energy}
\ee
where $\epsilon(n)$ is the energy density per particle, ${\bf v}$ the velocity field, $\mu$ the chemical potential and $n$ the density.

By terming $R_{\rm S}(\theta,\phi)$ the interface separating the superfluid from the normal component, the integral in Eq.(\ref{eq:energy}) splits in two parts. The internal superfluid core occupies the region $r<R_{\rm S}(\theta,\phi)$ and the surrounding normal phase is confined to $R_{\rm S}(\theta,\phi)<r<R_{\rm N}(\theta,\phi)$, where  $R_{\rm N}(\theta,\phi)$ is the Thomas-Fermi radius of the gas where the density vanishes.
The energy densities in the two phases are given, respectively, by 
\be
\epsilon_{\rm S}=\xi_{\rm S}\frac{3}{5}\frac{\hbar^2}{2m}(6\pi^2 n_{\rm S})^{{2}/{3}},
\label{eq:eS}
\ee
\be
\epsilon_{\rm N}=\xi_{\rm N}\frac{3}{5}\frac{\hbar^2}{2m}(6\pi^2 n_{\rm N})^{{2}/{3}},
\label{eq:eN}
\ee
where  $n_{\rm S}$ ($n_{\rm N}$) is the superfluid (normal) density and the dimensionless parameters $\xi_{\rm S}=0.42$ and $\xi_{\rm N}=0.56$ account for the role of interactions in the two phases. Their value has  been calculated in \cite{Carlsonxi,Stefanoxi} employing Quantum Monte Carlo simulations.

The equilibrium is found by minimizing the energy with respect to the densities and the velocity fields of the superfluid and normal part, as well as with respect to the position of the border surface.  Notice that this picture ignores surface tension effects, a plausible assumption in the limit of large samples. The superfluid velocity obeys the irrotationality constraint and can be written as ${\bf v_{\rm S}}=\nabla\Phi$.  Varying the energy with respect to the velocity potential $\Phi$ yields the continuity equation
\be
\nabla \cdot((\nabla \Phi -{\bf \Omega} \times {\bf r})n_{\rm S})=0,
\label{eq:graddens}
\ee
while variation with respect to the superfluid density $n_{\rm S}$ leads to 
\be
\mu=\xi_{\rm S}\frac{\hbar^2}{2m} (6 \pi^2 n_{\rm S})^{2/3}+V^{\rm S}({\bf r}),
\label{eq:locdensSF}
\ee
where $V^{\rm S}({\bf r})=V({\bf r})+\frac{1}{2}mv_{\rm S}^2-m\Omega({\bf r}\times{\bf v}_{\rm S})_{\rm Z}$ is the effective harmonic potential felt by the superfluid.  
%With the ansatz $\Phi=\alpha x y $ for the velocity field, the effective potential   is still harmonic with renormalized frequencies 
%$(\ox^{\rm S})^2=\ox^2+\alpha^2 -2\alpha\Omega$ and $(\oy^{\rm S})^2=\oy^2+\alpha^2 +2\alpha\Omega$.
%Using Eqs.(\ref{eq:graddens}) and (\ref{eq:locdensSF}) we find that $\alpha$ is a real root of
%\be
%2\alpha^3+\alpha (\ox^2+\oy^2-4\Omega^2)+\Omega(\ox^2-\oy^2)=0.
%\label{eq:alpha}
%\ee
%Eq.(\ref{eq:alpha}) predicts the occurrence of two branches. The first one (normal branch) starts at $\Omega=0$ and ends at $\Omega=\min\{\ox,\omega_y\}$, while the second (overcritical) branch starts at infinity, exhibits a backbending and ends at $\Omega=\max\{\omega_x,\oy\}$ (for more detailed discussion see \cite{alessio} and figures therein). 
 
Using the same procedure for the normal part without the irrotationality constraint we have $\bf{v_{\rm N}}={\bf\Omega}\times{\bf r}$, i.e. the normal component rotates rigidly.
The variation with respect to the normal density yields an equation similar to (\ref{eq:locdensSF})
\be
\mu=\xi_{\rm N}\frac{\hbar^2}{2m} (6 \pi^2 n_{\rm N})^{2/3}+V^{\rm N}({\bf r}),
\label{eq:locdensNM}
\ee
where the effective harmonic potential $V^{\rm N}({\bf r})$ is now quenched by the rigid rotation according to $(\ox^{\rm N})^2=\ox^2-\Omega^2$, $(\oy^{\rm N})^2=\oy^2-\Omega^2$.

By varying the energy (\ref{eq:energy}) with respect to $R_{\rm S}(\theta,\phi)$ we eventually find the equilibrium condition for the coexistence of the two phases in the trap.
The resulting equation implies that the pressure of the two phases be the same: $n_{\rm S}^2(\partial\epsilon_{\rm S}/\partial n_{\rm S})=n_{\rm N}^2(\partial\epsilon_{\rm N}/\partial n_{\rm N})$. Using Eqs.(\ref{eq:eS}) and (\ref{eq:eN}), one then predicts a density discontinuity at the interface given by
\be
\frac{n_{\rm N}}{n_{\rm S}}=\gamma\hspace{0.2cm}{\rm with} \hspace{0.2cm}\gamma=\left(\frac{\xi_{\rm S}}{\xi_{\rm N}}\right)^{{3}/{5}}=0.85 \; ,
\label{eq:jump}
\ee
independent of the angular velocity.
The equal pressure condition (\ref{eq:jump}) combined with Eqs.(\ref{eq:locdensSF}) and (\ref{eq:locdensNM}), results in the useful relationship   
\be 
(\mu-V^{\rm S}({\bf r}))=\gamma(\mu-V^{\rm N}({\bf r})),
\label{eq:muP0}
\ee
which determines the surface $R_{\rm S}(\theta,\phi)$ separating the superfluid and the normal part. 

In the following we assume $\omega_x=\omega_y\equiv \omega_\perp$ and we consider  the solution ${\bf v}_{\rm S}=0$,  corresponding to a non rotating axi-symmetric superfluid  and hence to $V^{\rm S}=V$. In this case we find 
%\be
%R_{\rm S}^2(\theta,\phi)=\frac{2\mu}{m}\frac{1-\gamma}{(1-\gamma)(\omega_z^2\cos^2\theta+\op^2
%\sin^2\theta)+\gamma\Omega^2 \sin^2\theta}
%\ee 
\be
R_{\rm S}^2(\theta)=\frac{2\mu}{m}\left(\omega_z^2\cos^2\theta+\op^2
\sin^2\theta+\frac{\gamma}{1-\gamma}\Omega^2 \sin^2\theta\right)^{-1},
\label{eq:RSF}
\ee 
where $\theta$ is the polar angle. On the other side the Thomas-Fermi radius $R_{\rm N}$ of the normal gas is fixed by the condition $\mu = V^{\rm N}$ yielding
\be
R_{\rm N}^2(\theta)=\frac{2\mu}{m}\left(\omega_z^2\cos^2\theta+\op^2
\sin^2\theta-\Omega^2\sin^2\theta\right)^{-1} \ge R_{\rm S}^2(\theta) \; .
\label{eq:RN}
\ee 
The value of $\mu$ is fixed by the normalization condition
\be
\int_{r<R_{\rm S} (\theta)} n_{\rm S}d{\bf r} + \int_{R_{\rm S}(\theta)<r<R_{\rm N}(\theta)} n_{\rm N}d{\bf r} = N  \; ,
\ee
where $N$ is the total number of particles. 
While for $\Omega=0$ the system is completely superfluid, for $\Omega>0$ it phase separates into a superfluid and a normal component characterized by the density jump (\ref{eq:jump}) at the interface. This behavior shares interesting analogies with the the phase separation between a superfluid and a normal component exhibited by polarized Fermi gases \cite{MIT2,Rice,MIT3} where a jump in the density at the interface is also predicted to occur \cite{normalLobo}. By tomographic techniques \cite{MIT3} it is nowadays possible to measure the density "in situ", thus the predicted discontinuity, and hence the value of $\xi_{\rm S}/\xi_{\rm N}$, should be observable experimentally. 

The radii $R_{\rm S}$ and $R_{\rm N}$ coincide at $\theta =0$ which means the  absence of the normal part along the $z$-axis. 
In the plane of rotation ($\theta=\pi/2$) the difference between the two radii is instead maximum and becomes larger and larger as $\Omega$ increases. In particular the radius of the superfluid is always smaller than the radius of the cloud in the absence of rotation, while the radius of the normal gas is always larger due to the bulge effect produced by the rotation.

\begin{figure}[htb]
\begin{center}
\includegraphics[height=5cm] {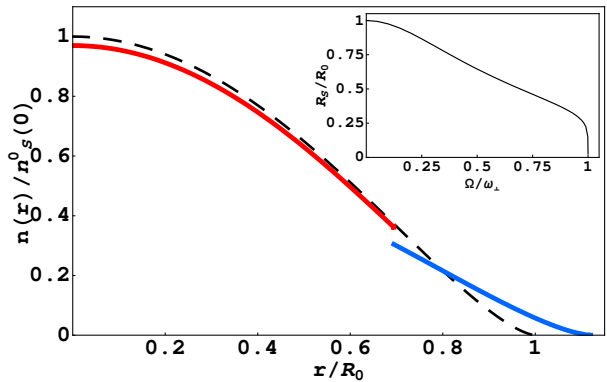}
\end{center}
\caption{Density profile for $\theta=\pi/2$ of the rotating Fermi gas at $\Omega=0.45 \omega_\perp$ (full line). The profile in the absence of rotation is also shown (dashed line). Inset: Superfluid radius versus $\Omega/\op$.}
\label{fig:density}
\end{figure}
In Fig.(\ref{fig:density}) we plot the densities $n_{\rm S}$ and $n_{\rm N}$ as a function of the radial coordinate at $\theta=\pi/2$ in a spherical trap for $\Omega=0.45\op$. The densities and the radial coordinate have been renormalized with respect to the central density $n_{\rm S}^0$ and the Thomas-Fermi radius $R_0$ of the superfluid at rest. The inset shows the  superfluid radius $R_{\rm S}$ renormalized  by $R_0$ as a function of the angular velocity $\Omega/\op$.

\begin{figure}[htb]
\begin{center}
\includegraphics[height=5cm] {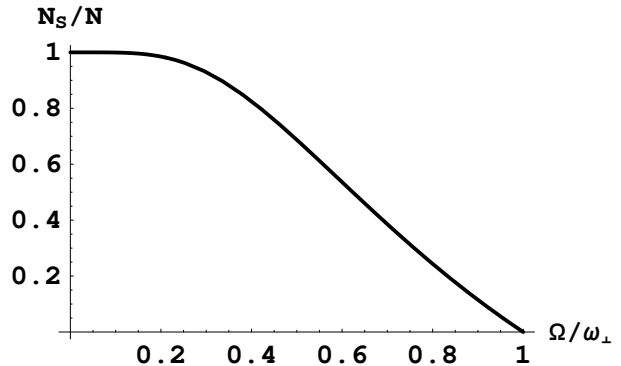}
\end{center}
\caption{Depletion of the superfluid as a function of the trap angular velocity for an axi-symmetric configuration (${\bf v}_{\rm S}$.}
\label{fig:NS}
\end{figure}

From the knowledge of the density profiles and from the radii Eqs.(\ref{eq:RSF}) and (\ref{eq:RN}) we can  calculate the number of particles in each phase. In Fig.(\ref{fig:NS}) we show the ratio between the number of particles $N_{\rm S}$ in the superfluid phase and the total number $N$ as function of the angular velocity. The higher the angular velocity, the more particles prefer to stay in the normal phase and thus the superfluid is depleted. At small angular velocities the depletion of the superfluid follows the law $N_{\rm S}/N=1-(\frac{\gamma}{1-\gamma})^{5/2}\ \Omega^5$.

\begin{figure}[htb]
\begin{center}
\includegraphics[height=5cm] {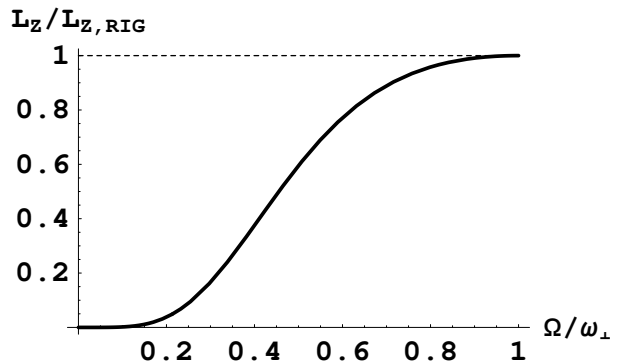}
\end{center}
\caption{Angular momentum in units of the rigid value as a function of the trap angular velocity for an axi-symmetric configuration.}
\label{fig:L}
\end{figure}

Another important observable is the angular momentum $L_{\rm Z}$. For an axi-symmetric configuration the superfluid does not carry angular momentum which is then provided only by the normal component: $L_{\rm Z}=\Omega\int d{\bf r}\  (x^2+y^2)n_{\rm N} $. The total angular momentum then increases with $\Omega$  and eventually reaches the rigid body value at $\Omega=\omega_\perp$ (see Fig.(\ref{fig:L})). The angular momentum of a rotating configuration has been measured in BEC's by studying the precession phenomena exhibited by the surface excitations \cite{Daliangmom}. 

While for small values of the angular velocity ($\Omega < 0.2\op$) the superfluid is robust, it is remarkable that even at angular velocities far from the centrifugal limit the depletion of the superfluid and hence the angular momentum of the system are sizable.

A major issue concerns the conditions of stability of the rotating configuration discussed in this work. Let us first consider the question of  energetic stability. We have shown that, in the frame rotating with the trap, the phase separated configuration  is energetically favoured with respect to the configuration where the whole gas is at rest and superfluid. This is true for any value of the angular velocity. When $\Omega$ exceeds a critical value of order $(\hbar/mR^2)\ln(R/d)$, where $d$ is the healing length fixed, at unitarity by the interparticle distance, quantized vortices become an even more favourable configuration. This energetic instability is not however expected to be a severe difficulty if one increases the angular velocity in an adiabatic way because the presence of a barrier inhibits  the access to the vortical configuration  as proven experimentally in the case of BEC's \cite{Dalibard3}. At higher angular velocities the axi-symmetric superfluid configuration is eventually expected to exhibit a surface energetic instability,  undergoing   a continuous shape deformation, similarly to what happens in Bose-Einstein condensates \cite{alessio,Dalibard1,Dalfovo}. This effect is accounted for by the solution ${\bf v}_{\rm S} \ne 0$ of  Eq.(\ref{eq:graddens}), corresponding to a spontaneous breaking of rotational symmetry in the superfluid component. 
In the case of Bose-Einstein condensates a quadrupole instability takes place at $\Omega_{cr}= \omega_\perp/\sqrt 2$ and for larger values of $\Omega$ the solution ${\bf v}_{\rm S}=0$ corresponds to the so called overcritical branch \cite{alessio}. A different value of $\Omega_{cr}$ is  predicted for the rotating Fermi gas, due to the new boundary condition imposed by the presence of the normal component. This condition requires that the velocity of the superfluid be tangential to the interface. In the simplest case of a 2D configuration we find the value $\Omega_{cr}=0.45 \omega_{\perp}$ for the emergence of a spontaneous quadrupole deformation. The experimental measurement of $\Omega_{cr}$ would provide a crucial test of the consequences of the phase separation caused by the rotation of the unitary Fermi gas. 

An even more challenging question concerns the emergence of dynamic instabilities. In the case of a rotating BEC a dynamic instability takes place at values of $\Omega$ slightly larger than $\omega_\perp/\sqrt2$ and corresponds to the appearance of imaginary components in the frequency of some hydrodynamic modes \cite{castin}. In the case  of the rotating Fermi gas discussed in this work a  dynamic instability might be associated with the Kelvin-Helmholtz instability of the interface between  two fluids in relative motion (see.e.g \cite{LL6}).
However, if the densities of the two fluids are different, an external force  stabilizes the two-fluid system against the appearance of complex frequencies in the low energy excitations of the interface \cite{LL6}. This is actually  our case where the density of the two phases exhibits the gap (\ref{eq:jump}) and the system  feels the external force produced by the harmonic confinement. We consequently expect that the system be dynamically stable at least for  moderately small values of the angular velocity.

In conclusion we have shown that an ultracold Fermi gas at unitarity can separate into a superfluid and a normal component as a consequence of the adiabatic ramping of the trap rotation. The formation of the rotating normal component requires that the trap transfers angular momentum to the gas within experimentally accessible times. Its realization would open the unique possibility of exploring the Fermi liquid behaviour of a strongly interacting gas at zero temperature. Important effects to investigate are, for example, the zero sound nature of the collective oscillations and the behaviour of viscosity. 
The  detailed study of the energetic instability associated with the spontaneous breaking of rotational symmetry as well as the effects of the rotation on the polarized phase ($N_\ua\neq N_\da$) will be the object of a future work.

We acknowledge stimulating discussions with Frederic Chevy, Stefano Giorgini and Lev Pitaevskii. We also acknowledge support by the Ministero dell'Istruzione, dell'Universit\`a e della Ricerca (MIUR).

\end{document}